\documentclass[preprint,amsmath,amssymb,aps,a4]{revtex4-2}
\usepackage{graphicx}
\usepackage{bm}
\usepackage{txfonts}
\usepackage{hyperref}
\usepackage[english]{babel}
\usepackage[autostyle, english = american]{csquotes}
\MakeOuterQuote{"}
\date{\today}
\newcommand{\be}{\begin{eqnarray}}
	\newcommand{\ee}{\end{eqnarray}}
\newcommand{\bfq}{{\bf q}_{\perp}}
\newcommand{\bfR}{{\bf R}_{\perp}}
\newcommand{\bfz}{{\bf 0}_{\perp}}

\newcommand{\bfk}{{\bf k}_{\perp}}

\newcommand{\bfki}{{\bf k}_{\perp i}}
\newcommand{\bfb}{{\bf b}_{\perp}}

\newcommand{\bfP}{{\bf P}_{\perp}}

\newcommand{\bfp}{{\bf p}_{\perp}}

\newcommand{\bfpi}{{\bf p}_{\perp i}}
\newcommand{\bfpip}{{\bf p}_{\perp i}^\prime}

\newcommand{\Dp}{{\bf \Delta}_{\perp}}

\newcommand{\lipr}{{\lambda_{i}}^\prime}

\newcommand{\bfr}{{\bf r}_{\perp}}
\usepackage{xcolor}
\begin{document}
\title{Transverse distortion and single-spin asymmetries for low-lying octet baryons}
\author{Navpreet Kaur}
\email{knavpreet.hep@gmail.com}
\affiliation{Department of Physics, Dr. B. R. Ambedkar National Institute of Technology, Jalandhar 144008, India}
\author{Harleen Dahiya}
\email{dahiyah@nitj.ac.in}
\affiliation{Department of Physics, Dr. B. R. Ambedkar National Institute of Technology, Jalandhar 144008, India}
	
\date{\today}
\begin{abstract}
Being the fountainhead of transverse distortion and asymmetries in accordance to $``SSA=GPD \ast FSI\,"$, we study the spin flip generalized parton distribution. We demonstrate this transverse deformation by using the quark-scalar diquark model and unveil a comparison among the low-lying strange baryons for distinct feasible combinations of quark-diquark pairs. Final-state interaction originating the correlation of a target spin and a virtual photon to the generated pion plane in semi-inclusive deep inelastic scattering has also been analyzed.
\end{abstract}
\maketitle
%
%
\section{Introduction\label{secintro}}
Several groups within the hadron physics community are making significant progress in answering the questions related to the building blocks of our universe such as genesis of nucleon mass \cite{Ji:2021qgo, Lorce:2021xku, Wang:2019mza, Lorce:2017xzd}, spin \cite{Ji:2016jgn, Alexandrou:2020sml}, nucleon interactions \cite{Cruz-Torres:2019fum, West:2020tyo} and the tomographic imaging of constituent partons (quarks and gluons) inside nucleons \cite{Celiberto:2021zww, Burkardt:2006td, Bacchetta:2022crh, Barry:2023qqh} with the baseline of quantum chromodynamics (QCD). Three-dimensional (3D) mapping of an internal structure of these nucleons and nucleon like structures such as baryons can be unwound theoretically by employing distribution functions \cite{Meissner:2009ww}. Hadron structure study has been upgraded from one-dimensional (1D) structure function in terms of collinear parton distribution functions (PDFs) \cite{Maji:2016yqo} to the multi-dimensional portraits \cite{Diehl:2015uka} in terms of generalized parton distributions (GPDs) \cite{Boffi:2007yc} and transverse momentum-dependent distributions (TMDs) \cite{Pasquini:2008ax}. Experimentally, GPDs can be computed via deeply virtual Compton scattering (DVCS) \cite{Xie:2023xkz} which is similar to the traditional Compton scattering with the replacement of a real photon by a virtual photon to scatter off a target hadron $X$ with the production of a real photon. If a meson is produced instead of a real photon, then this scattering process is coined as deeply virtual meson production (DVMP) \cite{Goloskokov:2007nt, Brooks:2018uqk}. On the other hand, TMDs correlate the transverse momentum and the polarizations of the constituent partons and hadrons illustrating a unique perspective on 3D tomography of a hadron \cite{Bacchetta:2010uj}. These TMDs measurements can be made from the semi-inclusive deep inelastic scattering (SIDIS) experiments \cite{Collins:2004nx, Accardi:2017pmi}. 

GPDs are the matrix elements of the non-local light-cone operators \cite{Belitsky:2005qn}. An inherent physical picture of hadrons can be prevailed in both momentum as well as impact parameter space \cite{Kumar:2015fta, Kaur:2018ewq}. Chiral even GPDs for purely transverse momentum transfer, $H_{X}^{q}(x,0,t)$ and $E_{X}^{q}(x,0,t)$ have deep pockets and can be used to explore the Dirac and Pauli form factors \cite{Zhang:2016qqg}. In the forward limit, they can reduce to PDFs \cite{Diehl:2003ny}, charge distribution \cite{Kim:2008ghb}, magnetization densities \cite{Kumar:2014coa} etc. GPDs in impact parameter space illustrate the spatial distribution of partons in the plane transverse to the momentum of a fast moving hadron \cite{Burkardt:2002hr}. The spin flip matrix element of GPDs $E_{X}^{q}(x,0,t)$ carries substantial information about the asymmetries. The physical significance of this spin flip GPD in the infinite momentum frame can be understood in terms of the distorted partonic distributions of $q$ flavored quarks in the transverse plane when the target state is transversely polarized. The fountainhead of this distortion is the existence of a non zero spin flip GPD \cite{Burkardt:2001ni}. This mechanism predicts a left-right asymmetry that shows the presence of single-spin asymmetry (SSA) as it is connected to the GPDs by the relation $``SSA=GPD \ast FSI\,"$ as demonstrated in Ref. \cite{Burkardt:2003je} for a scalar diquark model where FSI corresponds to final-state interaction.

The experimental observation of HERMES \cite{HERMES:1999ryv} and SMC \cite{Bravar:1999rq} collaborations have shown a significant amount of correlation between the spin of a target proton and the virtual photon to the generated pion plane in SIDIS with large virtuality as large as $Q^{2}=6 \, GeV^{2}$. This observation inspired the theoretical physicists to unfurl the fundamental concepts behind this correlation. SSA is time reversal odd and arises in QCD via phase difference in the different amplitudes of spin. In order to observe the correlation of hadron target spin and the virtual photon to hadron production plane, there must be two hadron spin amplitudes $M[\gamma^{\ast} X (J^{z})\rightarrow F]$ with total spin $J^{z}= \, \Uparrow/\Downarrow$ which couple to the same final state, however, with different complex phases. These correlations are directly connected to the imaginary part of the product of these amplitudes as $(M[J^{z}= \, \Uparrow]^\ast M[J^{z}= \, \Downarrow])$. Therefore, to analyze this SSA, understanding of QCD at amplitude level is necessary. Both initial-state interaction (ISI) \cite{Boer:2002ju} as well as FSI \cite{Brodsky:2002pr} in gauge theory can be employed to demonstrate this mechanism. It has been shown in Ref. \cite{Brodsky:2002cx} that, at leading twist in perturbative QCD, the FSI with gluon exchange between an outgoing quark $q$ and the spectator $n$ leads to SSA in deep inelastic lepton-proton scattering.

GPDs, TMDs and the related  physical observables that can be extracted from these distributions have been extensively explored to scrutinize the internal structure of nucleons not only at leading twist but also for higher twists \cite{Mukherjee:2002xi, Dodson:2021rdq, sstwist4}. These nucleons have spin-parity quantum number, $J^{P}=\frac{1}{2}^{+}$ and in accordance to group theory, they belong to a baryon octet in which nucleons are the isospin companions with light up $u$ and down $d$ flavor of quarks. Other members of octet baryon such as $\Lambda$, isospin companions $\Sigma$ and $\Xi$ with same spin-parity quantum number have a comparatively heavier strange $s$ flavor of quark. Due to the presence of this $s$ flavor of quark, their lifetimes are very small and it becomes difficult to access them in the experimental scattering processes. In this present work, we will analyze the amount of transverse distortion and SSA corresponding to each constituent flavor of quarks in strange baryons to unveil its mass effect in the scalar diquark model.

A QCD inspired scalar diquark model of a hadron represents a simplistic view of an active quark and a spectator of diquark. Electromagnetic form factors have been studied in this model and the results are in good agreement with experimental data \cite{Hwang:2007tb}. Along with this, the distribution functions such as GPDs, charge densities \cite{Kim:2008ghb} and angular momentum \cite{Lorce:2017wkb} have also been investigated successfully with simple scalar diquark model. This model contains the tree and one-loop amplitudes required for SSA and have all the Lorentz symmetries as the light-front wavefunctions (LFWFs) attributed to quarks along with FSI are formulated perturbatively. Because of this feature, the representation of the Wilson line becomes crucial as the  phase factor in Wilson line explains the FSI in SIDIS \cite{Brodsky:2005yw}. Basically, Wilson line phase factor for an active quark demonstrates the phase factor of a propagator when an active quark leaves its parent hadron and this phase factor is not invariant under time-reversal as a consequence of which asymmetry pop up \cite{Burkardt:2003je}. We have adopted the scalar diquark model to present the qualitative analysis of trannsverse distortiona and SSA among low-lying octet baryons.

The present paper is organized as follows. Section \ref{secModDes} comprises the description of the scalar diquark model. All the necessary feeds required to do the calculations have been presented in Section \ref{secNumPara} followed by discussion over the transverse distortion of strange baryons in Section \ref{sectransdistort}. Further, comparative analyzes of the azimuthal transverse and longitudinal SSAs are illustrated and discussed in Section\ref{secSSA}. We conclude our outcomes in Section \ref{secCon}.
\section{Scalar Diquark model \label{secModDes}}
An invariant mass square is the eigenvalue of an invariant light-cone Hamiltonian,  $H_{LC}^{QCD}=P^{+} P^{-}-{\bf P_\perp^2}$ with a hadron as an eigenstate.  Here, $P^{+}$ and $\bfP$ are the momentum generators which are kinematical quantities as they do not vary with an interaction. On the other hand, $P^{-}$ is  the generator that governs the light-front time translations. The composition of a hadron can be described by its wavefunction in terms of the momenta and spin projection of the building blocks of the hadron. So, the eigensolution of a baryon, projected on its color singlet eigenstate $|\mathcal{N}\rangle$ of the free Hamiltonian $H_{LC}^{QCD}$ at fixed light-cone time $\tau=t+z/c$ \cite{Dirac:1949cp}, can be written as \cite{Brodsky:2000rt}
\be
|\psi_{X}(P^{+},{\bf P_\perp)}\rangle &=& \sum_{\mathcal{N}} \prod_{i=1}^{\mathcal{N}} \frac{dx~  d^2\bfki}{2(2\pi)^3\sqrt{x_{i}}} \, 16 \pi^{3} \, \delta \bigg(1-\sum_{i=1}^{\mathcal{N}} x_{i}\bigg) \, \delta^{(2)} \bigg(\sum_{i=1}^{\mathcal{N}}\bfki\bigg) \nonumber \\		
&\times& \psi_{\mathcal{N}}(x_{i},\bfki,\lambda_{i})|\mathcal{N}; x_{i} P^{+},x_{i}\bfP+\bfki,\lambda_{i}\rangle \, ,
\ee
where $\bfki$ and $\lambda_{i}$ symbolize the light-cone intrinsic transverse momentum and helicity carried by an \textit{i}th component of a hadron respectively. $x_{i}=k_{i}^{+}/P^{+}$ is its longitudinal light-front momentum fraction. We have adopted the light-cone gauge, $A^{+}=0$. The LFWFs $\psi_{\mathcal{N}/X}$ are free from the dependence of hadron's momenta, $P^{+}$ and $\bfP$ which project the hadron state on the Fock state $|\mathcal{N}\rangle$. The  LFWF of a hadron corresponding to each Fock-state  with total spin $J^{z}$ is represented by $\psi_{\mathcal{N}}^{J^{z} X}(x_{i},\bfki,\lambda_{i})$ with
\begin{equation}
	k_{i}=(k_{i}^{+},k_{i}^{-},\bfki)=\bigg(x_{i}P^{+},\frac{\bfki^{2}+m_{i}^{2}}{x_{i}P^{+}},\bfki\bigg) \, .
\end{equation}
The states are normalized as
\begin{equation}
	\langle \mathcal{N}; p_{i}^{\prime +}, \bfpip, \lambda_{i}^{\prime} |\mathcal{N};  p_{i}^{+},\bfpi,\lambda_{i}\rangle = \prod_{i=1}^{\mathcal{N}} 16 \pi^{3} \, p_{i}^{+} \, \delta(p_{i}^{\prime +}-p_{i}^{+}) \, \delta^{(2)} (\bfpip-\bfpi) \, {\delta}_{\lipr \lambda_{i}} \, .
\end{equation}
Based on one-loop quantum fluctuations of the Yukawa theory \cite{Brodsky:2000ii},  there are two possible spin combinations for the two particle Fock state. For a hadron with $J^{z}=\,\Uparrow$ it can be  written as
\be
|\psi_{2 particle}^{\Uparrow X}(P^{+},\bfP=\bf{0}_{\perp})\rangle & =& \int \frac{dx~  d^2\bfki}{2(2\pi)^3\sqrt{x(1-x)}} \, \bigg[\psi_{+\frac{1}{2}}^{\Uparrow X}(x,\bfk)\bigg|+\frac{1}{2};x P^{+},\bfk\bigg\rangle \nonumber \\
&+& \bigg[\psi_{-\frac{1}{2}}^{\Uparrow X}(x,\bfk)\bigg|-\frac{1}{2};x P^{+},\bfk\bigg\rangle \bigg] \, ,
\ee
where
\be
\psi_{+\frac{1}{2}}^{\Uparrow X}(x,\bfk) &=& \bigg(M_{X}+\frac{m_{q}}{x}\bigg) \, \varphi_{X} \, , \nonumber \\
\psi_{-\frac{1}{2}}^{\Uparrow X}(x,\bfk) &=& -\frac{(k^{1}+\iota k^{2})}{x} \, \varphi_{X}\, .
\label{WFSPH}
\ee
Similarly, corresponding to $J^{z}=\,\Downarrow$ hadron, the two particle Fock state is given by
\be
|\psi_{2 particle}^{\Downarrow X}(P^{+},\bfP=\bf{0}_{\perp})\rangle & =& \int \frac{dx~  d^2\bfki}{2(2\pi)^3\sqrt{x(1-x)}} \, \bigg[\psi_{+\frac{1}{2}}^{\Downarrow X}(x,\bfk)\bigg|+\frac{1}{2};x P^{+},\bfk\bigg\rangle \nonumber \\
&+& \bigg[\psi_{-\frac{1}{2}}^{\Downarrow X}(x,\bfk)\bigg|-\frac{1}{2};x P^{+},\bfk\bigg\rangle \bigg] \, ,
\ee
where
\be
\psi_{+\frac{1}{2}}^{\Downarrow X}(x,\bfk) &=& \frac{(k^{1}-\iota k^{2})}{x} \, \varphi_{X} \, , \nonumber \\
\psi_{-\frac{1}{2}}^{\Downarrow X}(x,\bfk) &=& \bigg(M_{X}+\frac{m_{q}}{x}\bigg) \, \varphi_{X} \, .
\label{WFSNH}
\ee
The scalar part $\varphi$ has the form
\be
\varphi_{X}=\varphi_{X}(x,\bfk)=\frac{\frac{g}{\sqrt{1-x}}}{M^{2}_{X}-\frac{\bfk^{2}+m^{2}_{q}}{x}-\frac{\bfk^{2}+\mu^{2}_{n}}{1-x}} \, .	\label{scalar}
\ee
With the help of Fourier transformation, momentum space can be swapped to the impact parameter space as
\be
\psi(x,\bfb) = \frac{1}{1-x} \int \frac{d^2 \bfk}{4\pi^{2}} e^{\frac{i \, \bfk \cdot \bfb}{1-x}} \psi (x,\bfk) \, ,
\ee
where $\bfb=b_{\perp}(\cos \phi_{b}, \,\sin \phi_{b})$.
By employing this transformation, we have \cite{Lorce:2017wkb}
\be
\psi_{+\frac{1}{2}}^{\Uparrow X}(x,\bfb) &=& -\frac{g}{2 \pi} \frac{x M_{X}+m_{q}}{\sqrt{1-x}} K_{0}(Z) \, ,  \\
\psi_{-\frac{1}{2}}^{\Uparrow X}(x,\bfb) &=& \frac{ig}{2 \pi} \frac {\sqrt{\mathcal{M}_{X}^{un}(x)} e^{i \phi_{b}}}{\sqrt{1-x}} K_{1}(Z) \, , \\
\psi_{+\frac{1}{2}}^{\Downarrow X}(x,\bfb) &=& \frac{-ig}{2 \pi} \frac {\sqrt{\mathcal{M}_{X}^{un}(x)} e^{-i \phi_{b}}}{\sqrt{1-x}} K_{1}(Z),  \\
\psi_{-\frac{1}{2}}^{\Downarrow X}(x,\bfb) &=&  -\frac{g}{2 \pi} \frac{x M_{X}+m_{q}}{\sqrt{1-x}}  K_{0}(Z) \, ,
\ee
where
\be
Z=\frac{\sqrt{\mathcal{M}_{X}^{un} (x)} |\bfb|}{1-x} \, ,
\ee
\be
\mathcal{M}_{X}^{un} (x) &=& m_{q}^{2}(1-x)+\mu^{2}_{n}x-M^{2}_{X}x(1-x) \, ,
\ee
and $K_{p}$ is the $p$th order modified Bessel function of the second kind.
\section{Numerical parameters \label{secNumPara}}
The calculations performed to study the transverse distortion and SSA have $M_{X}$, $m_{q}$ and $\mu_{n}$ as the input parameters which respectively denote the masses of baryons, quarks and diquarks under consideration. Following the particle data group and Ref. \cite{Zhang:2016qqg}, we have tabulated the values of these parameters in Table \ref{tab_bmass} and \ref{tab_qmass}.
\begin{table}[h]
\centering
\begin{tabular}{|c|c|c|c|c|c|}
	\hline
	$\text{Particle $(X)$}  $~~&~~$ \Lambda $~~&~~$ \Sigma^{+} $~~&~~$ \Sigma^{o} $~~&~~$ \Xi^{o} $ \\
	\hline
	$\text{Mass, $M_{X}$ ($GeV$)} $~~&~~$ 1.115 $~~&~~$ 1.189 $~~&~~$ 1.192 $~~&~~$ 1.314 $ \\
	\hline
\end{tabular}
\caption{Masses of strange baryons used in the present calculations.}
\label{tab_bmass} 
\end{table}
\begin{table}[h]
\centering
\begin{tabular}{|c|c|c|c|c|c|}
	\hline
	$\text{Quark flavor}  $~~&~~$ u/d  $~~&~~$  s $~~&~~$ uu/ud $~~&~~$ us/ds $~~&~~$ ss $ \\
	\hline
	$\text{Mass ($GeV$)} $~~&~~$ 0.33 $~~&~~$ 0.48 $~~&~~$ 0.80 $~~&~~$ 0.95 $~~&~~$ 1.10 $ \\
	\hline
\end{tabular}
\caption{Masses of quark flavors and their combinations used in the present calculations.}
\label{tab_qmass} 
\end{table}
Along with these input parameters, we have $C_{F}$ and $\alpha_{s}$ which symbolize the color factor and coupling constant respectively. Their values are considered to be $\frac{4}{3}$ and $0.3$ respectively \cite{Brodsky:2002cx}. Contour plots of transverse distortion are in the units of $\frac{g^{2}}{4 \pi}$. For convenience, we have treated $u$ and $d$ quarks alike as a result of which outcomes for $\Sigma^{+}$ and $\Sigma^{o}$ do not show noteworthy dissimilarities and $\Sigma$ (=$\Sigma^{+}$=$\Sigma^{o}$). 
\section{Transverse distortion of the parton distributions \label{sectransdistort}}
\begin{figure*}
	\centering
	\begin{minipage}[c]{0.98\textwidth}
		(a)\includegraphics[width=5.7cm]{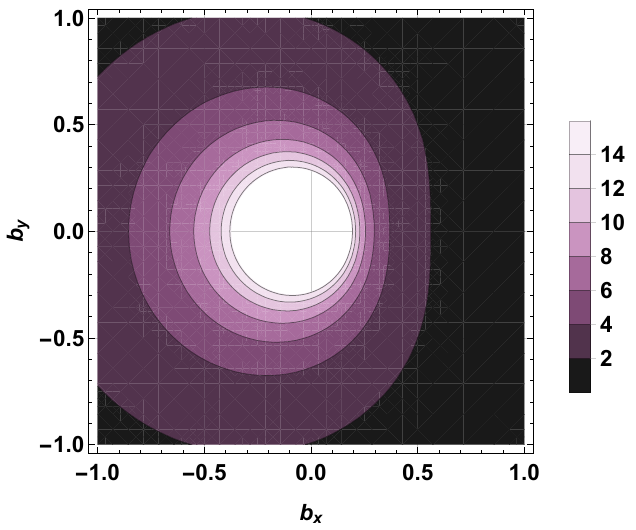}
		\hspace{0.03cm}
		(b)\includegraphics[width=5.7cm]{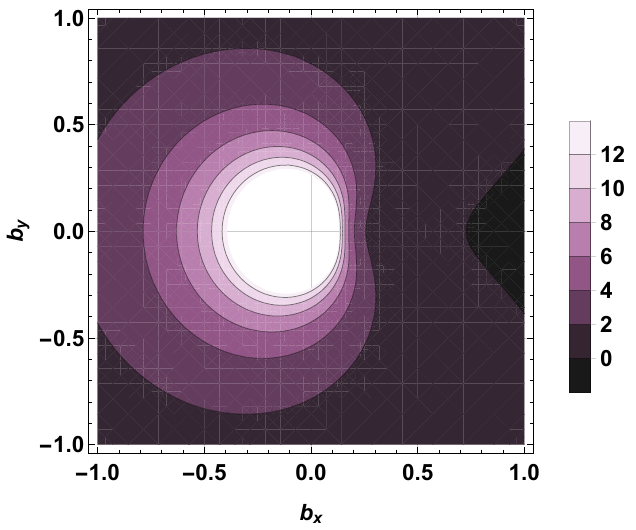}
		\hspace{0.03cm}
		(c)\includegraphics[width=5.7cm]{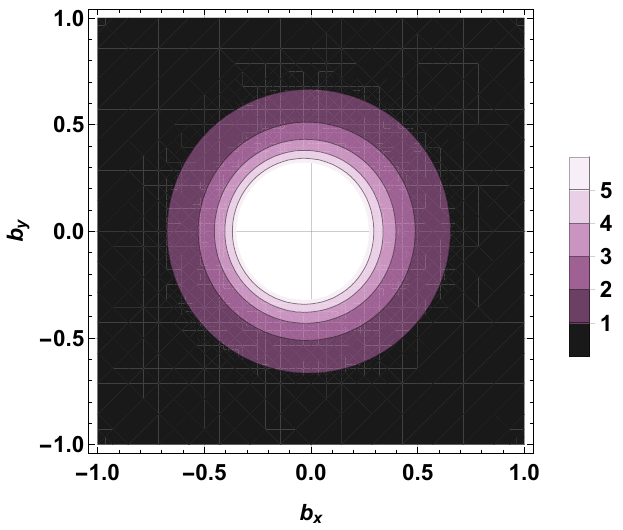}
		\hspace{0.03cm}
		(d)\includegraphics[width=5.7cm]{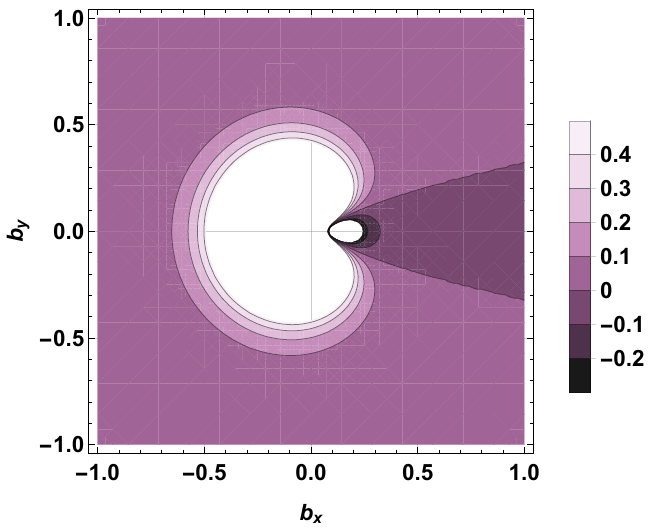}
		\hspace{0.03cm}
	\end{minipage}
	\caption{\label{Fig1LamU} (Color online)  Transverse distortion for $u$ quark flavor in $\Lambda$ for longitudinal momentum fraction $ x=0.2, 0.4, 0.6$ and $0.8$ respectively.}
\end{figure*}

\begin{figure*}
	\centering
	\begin{minipage}[c]{0.98\textwidth}
		(a)\includegraphics[width=5.7cm]{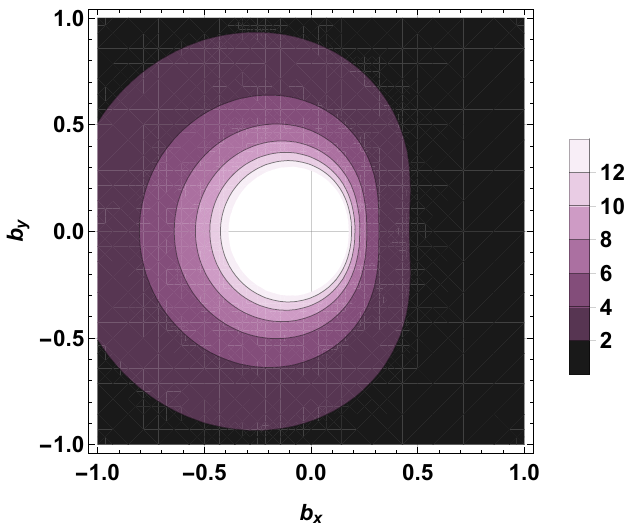}
		\hspace{0.03cm}
		(b)\includegraphics[width=5.7cm]{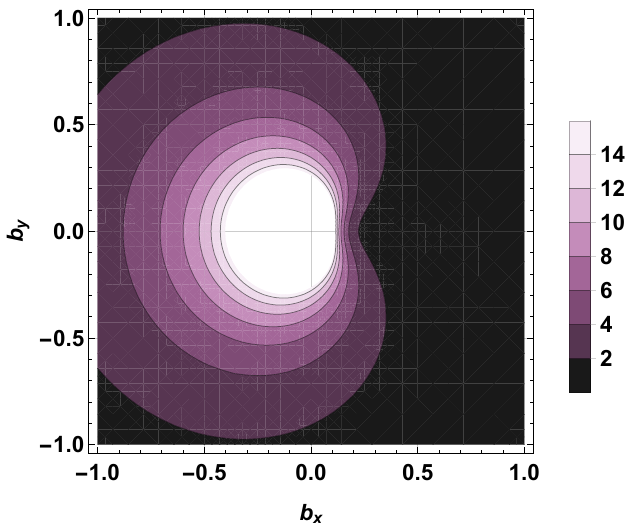}
		\hspace{0.03cm} \\
		(c)\includegraphics[width=5.7cm]{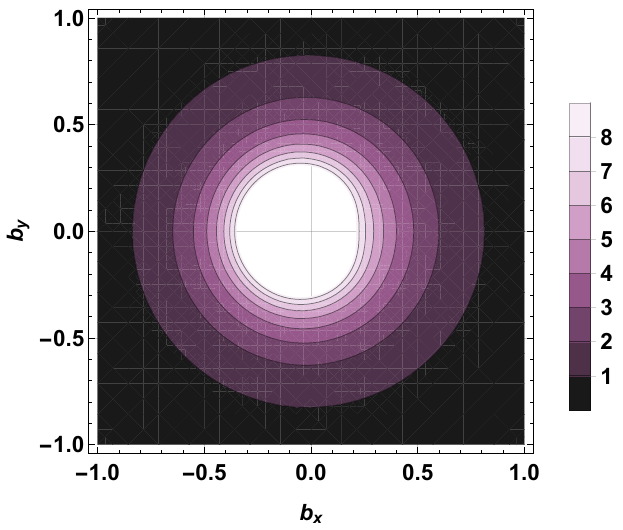}
		\hspace{0.03cm}
		(d)\includegraphics[width=5.7cm]{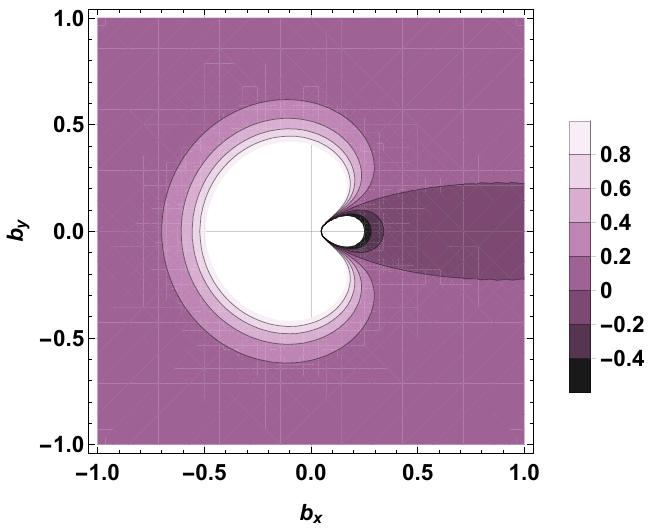}
		\hspace{0.03cm}
	\end{minipage}
	\caption{\label{Fig2LamS} (Color online) Transverse distortion for $s$ quark flavor in $\Lambda$ for longitudinal momentum fraction $ x=0.2, 0.4, 0.6$ and $0.8$ respectively.}
\end{figure*}

\begin{figure*}
	\centering
	\begin{minipage}[c]{0.98\textwidth}
		(a)\includegraphics[width=5.7cm]{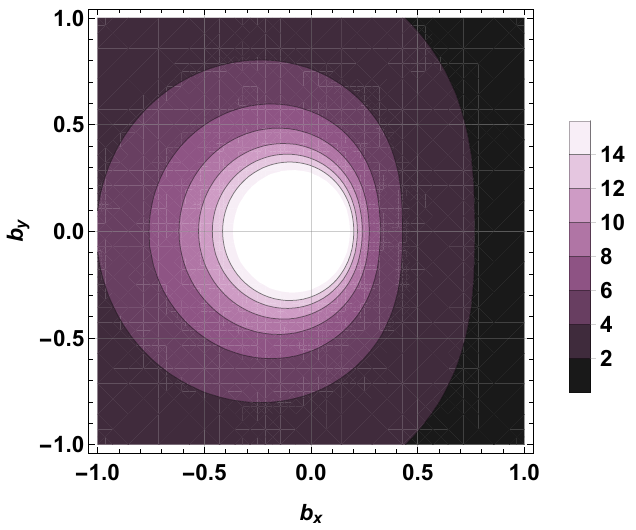}
		\hspace{0.03cm}
		(b)\includegraphics[width=5.7cm]{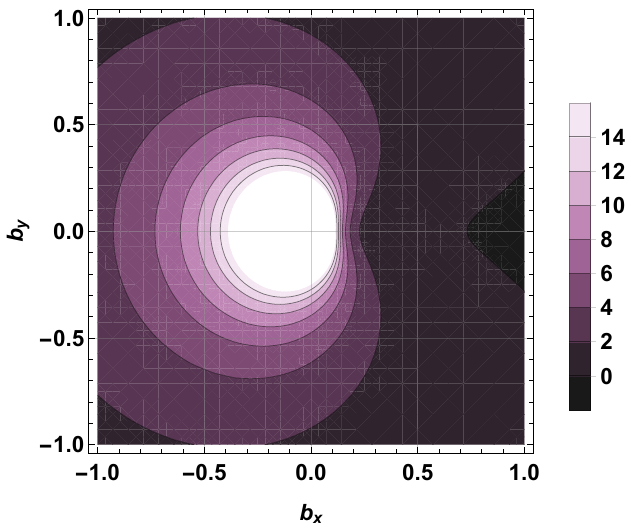}
		\hspace{0.03cm}\\
		(c)\includegraphics[width=5.7cm]{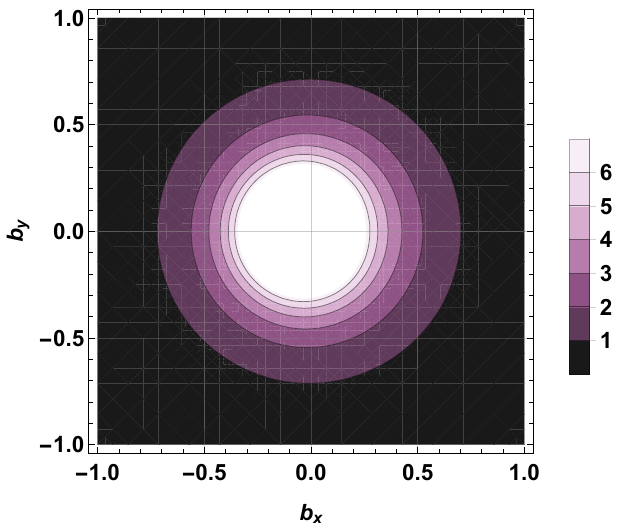}
		\hspace{0.03cm}
		(d)\includegraphics[width=5.7cm]{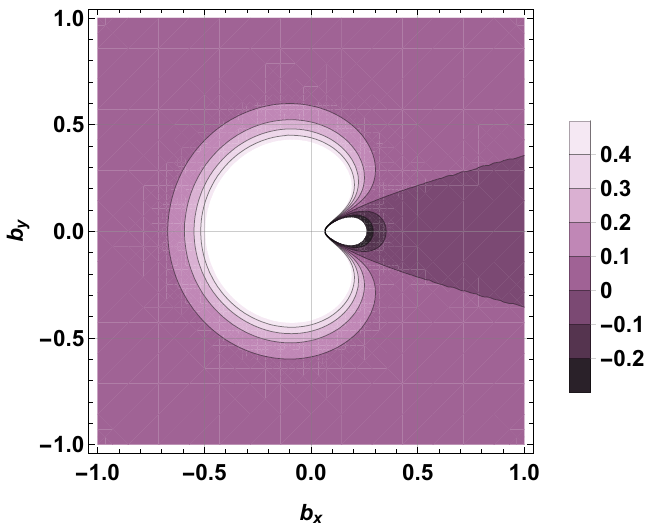}
		\hspace{0.03cm}
	\end{minipage}
	\caption{\label{Fig3Consu} (Color online)  Transverse distortion for $u$ quark flavor in $\Sigma^{+}$ for longitudinal momentum fraction $ x=0.2, 0.4, 0.6$ and $0.8$ respectively.}
\end{figure*}

\begin{figure*}
	\centering
	\begin{minipage}[c]{0.98\textwidth}
		(a)\includegraphics[width=5.7cm]{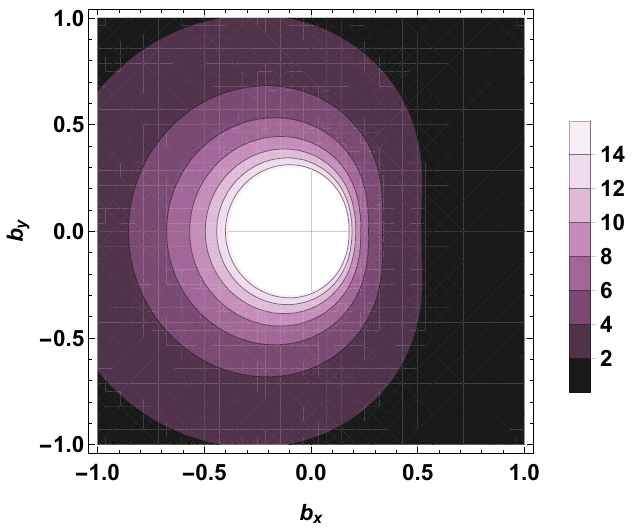}
		\hspace{0.03cm}
		(b)\includegraphics[width=5.7cm]{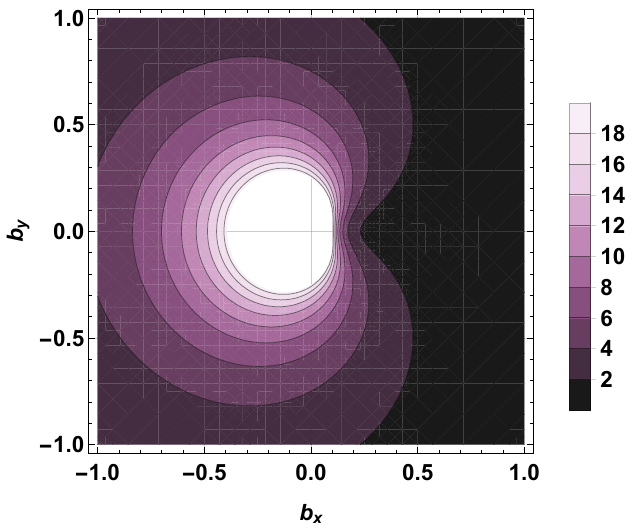}
		\hspace{0.03cm} \\
		(c)\includegraphics[width=5.7cm]{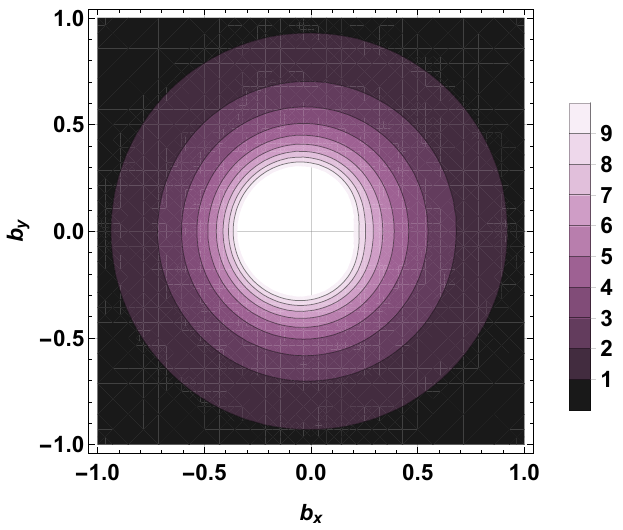}
		\hspace{0.03cm}
		(d)\includegraphics[width=5.7cm]{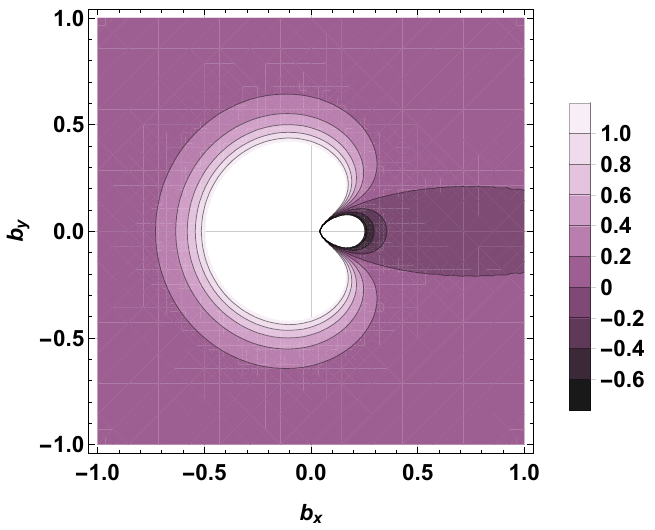}
		\hspace{0.03cm}
	\end{minipage}
	\caption{\label{Fig4Conss} (Color online) Transverse distortion for $s$ quark flavor in $\Sigma^{+}$ for longitudinal momentum fraction $ x=0.2, 0.4, 0.6$ and $0.8$ respectively.}
\end{figure*}

\begin{figure*}
	\centering
	\begin{minipage}[c]{0.98\textwidth}
		(a)\includegraphics[width=5.7cm]{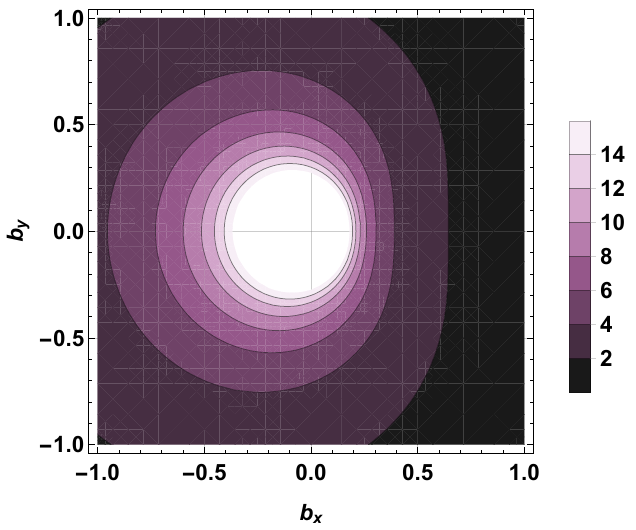}
		\hspace{0.03cm}
		(b)\includegraphics[width=5.7cm]{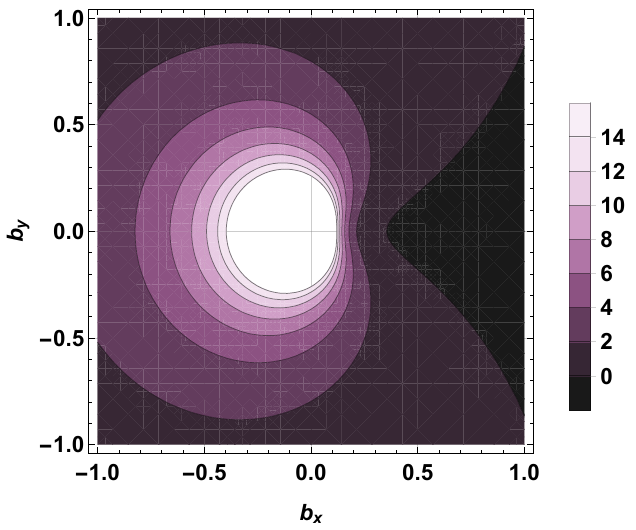}
		\hspace{0.03cm} \\
		(c)\includegraphics[width=5.7cm]{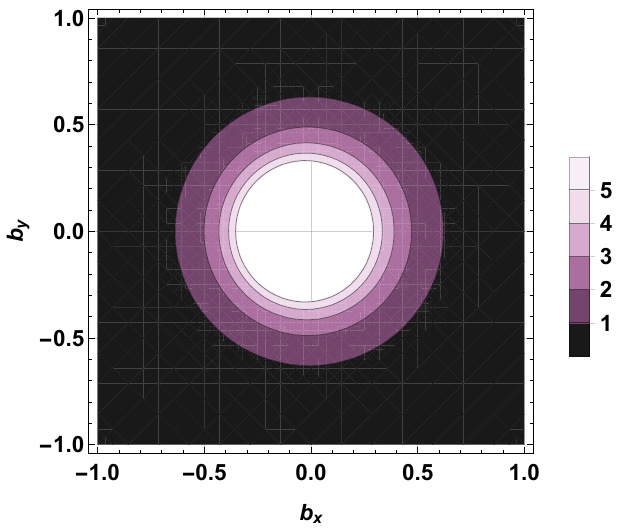}
		\hspace{0.03cm}
		(d)\includegraphics[width=5.7cm]{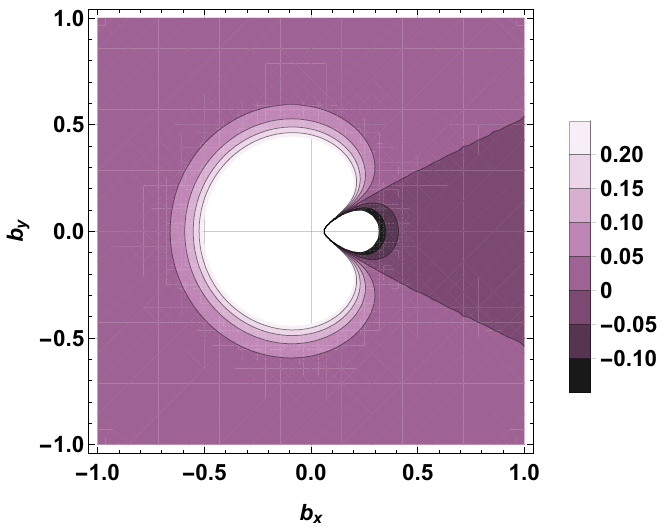}
		\hspace{0.03cm}
	\end{minipage}
	\caption{\label{Fig5Conxu} (Color online) Transverse distortion for $u$ quark flavor in $\Xi^{o}$ for longitudinal momentum fraction $ x=0.2, 0.4, 0.6$ and $0.8$ respectively.}
\end{figure*}

\begin{figure*}
	\centering
	\begin{minipage}[c]{0.98\textwidth}
		(a)\includegraphics[width=5.7cm]{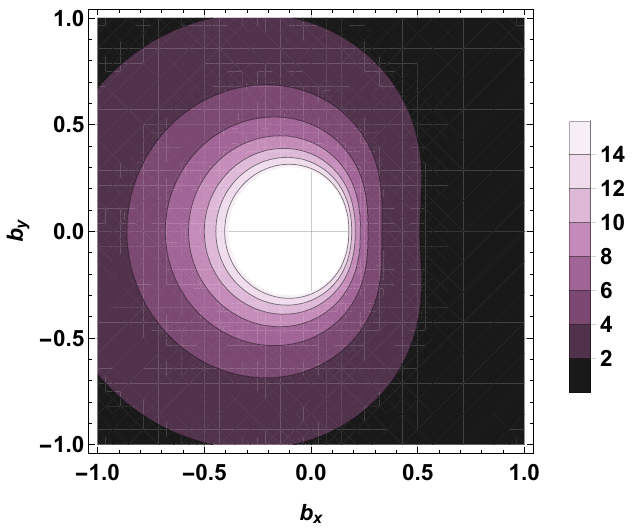}
		\hspace{0.03cm}
		(b)\includegraphics[width=5.7cm]{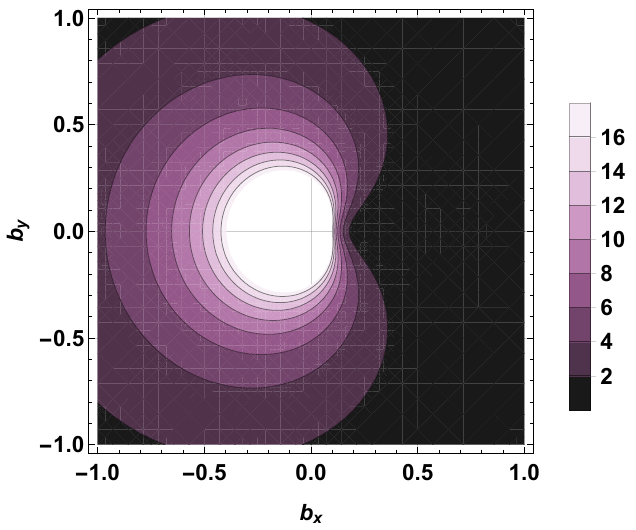}
		\hspace{0.03cm} \\
		(c)\includegraphics[width=5.7cm]{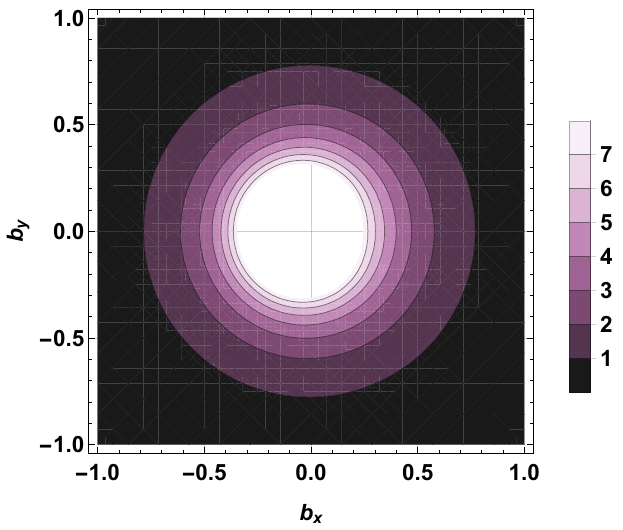}
		\hspace{0.03cm}
		(d)\includegraphics[width=5.7cm]{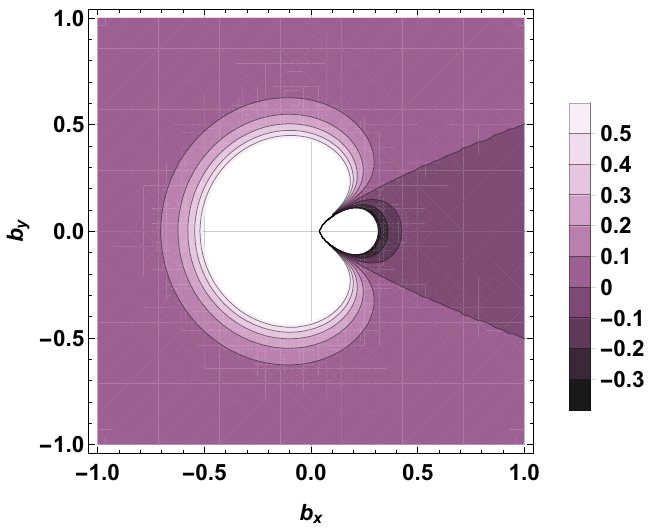}
		\hspace{0.03cm}
	\end{minipage}
	\caption{\label{Fig6Conxs} (Color online) Transverse distortion for $s$ quark flavor in $\Xi^{o}$ for longitudinal momentum fraction $ x=0.2, 0.4, 0.6$ and $0.8$ respectively.}
\end{figure*}

A comprehensive illustration of the microscopic structure of a baryon in terms of its constituent quarks can be obtained via GPDs. The distributions of GPDs are categorized as chiral even and chiral odd based on chiral symmetry concept. Among chiral even GPDs, there are $H_{X}^{q}(x,0,t)$ and $E_{X}^{q}(x,0,t)$ GPDs which do not depend on an active quark helicity. GPD $H_{X}^{q}(x,0,t)$ conserves the baryon helicity, describes the distribution of an unpolarized quarks inside an unpolarized baryon and is labeled as $``$unpolarized GPD$"$. Whereas GPD $E_{X}^{q}(x,0,t)$ involves  baryon flip helicity, description of a transversely polarized quark distribution inside an unpolarized baryon and is labeled as $``$transversely polarized GPD$"$. These quark helicity independent, chiral even GPDs can be obtained from the matrix elements of the bilinear vector currents as \cite{Brodsky:2000xy}
\be
\frac{1}{2}\int \frac{dy^{-}}{2 \pi}&e^{ix P^{+}y^{-}}& \bigg\langle P^{\prime}\bigg|\bar{\psi}\bigg(\frac{-y}{2}\bigg) \,\gamma^{+} \,\psi\bigg(\frac{y}{2}\bigg)\bigg|P\bigg\rangle \bigg|_{y^{+}=0, \bf{y_{\perp}}=0}  \nonumber \\
&=&\frac{1}{2 \bar{P^+}}\bar{u}(P^\prime)\bigg[H_{X}^{q}(x,0,t) \,\gamma^{+}+E_{X}^{q}(x,0,t) \,\frac{i \sigma^{+\alpha}(-\Delta_{\alpha})}{2M_{X}}\bigg]u(P) \, .
\label{GPDcorr}
\ee 
Here, $\bar{P}$ denotes the average momentum of the initial and final state of a baryon with $u(P)$ and $\bar{u}(P^\prime)$ as the light-cone spinors of the initial and final baryons. 
For a transversely polarized target, whenever a spin-flip GPD exists, a distortion in the parton distributions of quarks is observed. To develop an interpretation of these distortions, one has to consider amplitudes where the baryon and/or quark helicity flips. Therefore, we have considered $E^{q}_{X}(x,\zeta=0,t)$ which corresponds to the helicity flip amplitude of a baryon. To investigate the density interpretation of the $E^{q}_{X}(x,\zeta=0,t)$, consider a state that is polarized in the $y$-direction in the infinite momentum frame. We have
\be
|Y\rangle = \frac{1}{\sqrt{2}} \big[|p^{+},\bfR=\bfz,\uparrow\rangle + i \, |p^{+},\bfR=\bfz,\downarrow\rangle \big] \, ,
\ee
where
\be
|p^{+},\bfR=\bfz,\lambda\rangle = N \int d^{2} \bfp |p^{+},\bfp,\lambda\rangle \, .
\ee
By employing the Fourier transformation on GPDs, obtained from Eq. (\ref{GPDcorr}), the unpolarized quark distribution for the above defined state can be written in terms of the impact parameter space coordinates as 
\be
q_{\hat{y}}^{X_{q}} (x,\bfb) &=& \int \frac{d^{2} \Dp }{(2 \pi)^{2}} e^{i \Dp \cdot \bfb} \bigg[ H_{X}^{q}(x,0,t) + i \frac{\Dp^{x}}{2M_{X}} E_{X}^{q}(x,0,t) \bigg] \nonumber \\
&=&\mathcal{H}_{X}^{q} (x,\bfb)+ \frac{1}{2M_{X}} \frac{\partial}{\partial b^{x}} \mathcal{E}_{X}^{q} (x,\bfb) \, .
\label{transdistort}
\ee
Using the LFWFs expressed in the impact parameter space, the unpolaized chiral-even GPDs can be written as follows
\be 
\mathcal{H}_{X}^{q} (x,\bfb) &=& \frac{1}{2(2\pi)} \bigg[ \big| \psi^{\uparrow X}_{+\frac{1}{2}} (x,\bfb) \big|^{2} + \big| \psi^{\uparrow X}_{-\frac{1}{2}}  (x,\bfb) \big||^{2} \bigg]\, , \\
-\frac{1}{2} \bigg(i \frac{\partial}{\partial b^{x}} + \frac{\partial}{\partial b^{y}}\bigg) \, \mathcal{E}_{X}^{q} (x,\bfb)&=& \frac{1}{2(2\pi)} \bigg[  \psi^{\uparrow \ast X}_{+\frac{1}{2}} (x,\bfb) \, \psi^{\downarrow X}_{+\frac{1}{2}} (x,\bfb)  + \psi^{\uparrow \ast X}_{-\frac{1}{2}}  (x,\bfb) \, \psi^{\downarrow X}_{-\frac{1}{2}}  (x,\bfb)  \bigg]\, .
\ee 
The explicit expression of these GPDs, on substituting the values of LFWFs, come out to be 
\be 
\mathcal{H}_{X}^{q} (x,\bfb) &=& \frac{g^{2}}{16 \pi^{3}(1-x)} \bigg[(x M_{X}+m_{q})^{2} [K_{o}(Z)]^{2} + \mathcal{M}_{X}^{un}(x) [K_{1}(Z)]^{2}\bigg]\, , \\
\mathcal{E}_{X}^{q} (x,\bfb) &=& \frac{g^{2}}{16 \pi^{3}} 2 M_{X} (x M_{X}+m_{q})^{2} [K_{o}(Z)]^{2} \, .
\ee
Transverse distortion observed for each constituent flavors of quark in strange baryons are presented in Figs. \ref{Fig1LamU}-\ref{Fig6Conxs} for different values of $x$ $(0.2, 0.4, 0.6 $ and $0.8)$. In context to corporeal picture of Eq. (\ref{transdistort}), left side distortion in the parton distribution has been observed for a transversely polarized baryon moving with high momentum. Both GPDs $\mathcal{H}^{q}_{X}(x,\bfb)$ and  $\mathcal{E}^{q}_{X}(x,\bfb)$ are positive smooth functions of $|\bfb|$. The $b^{x}$-derivative of the positive function of $\mathcal{E}^{q}_{X}(x,\bfb)$ gives a positive value for negative $b^{x}$ and a negative value for positive $b^{x}$. Therefore, on adding $\mathcal{H}^{q}_{X}(x,\bfb)$ to $b^{x}$-derivative of $\mathcal{E}^{q}_{X}(x,\bfb)$, shifting effect in the parton distribution is obvious. This shifting effect can also be viewed from a semi-classical window in which the superposition of translatory and orbital motion of the quarks gives rise to the distortion when a baryon is transversely polarized with respect to the motion of a baryon. For a baryon accelerated in the $z$-direction, spin and orbital angular momentum are parallel to each other but directed in perpendicular direction with respect to motion of a baryon. This orientation adds up an orbital motion to the momentum towards the right side of the baryon whereas it gets subtracted from the left side. It implies that quarks on the right side get elevated to large momentum fraction $x$ whereas left side quarks deaccelerated to small momentum fraction $x$. Acceleration and deacceleration gets reversed if the spin and orbital angular momentum become anti-parallel to each other. In our case, as the shifting for all the flavors of quark is towards the left side, it corresponds to a situation in which spin and orbital angular momentum are anti-parallel to each other. \par

The transverse distortion corresponding to each flavor of quark portrayed in Figs. \ref{Fig1LamU}-\ref{Fig6Conxs} shows that the distortion for an active $u$ flavored quark spreads out on moving from momentum fraction $x=0.2$ to $0.4$ and thereafter shrinks down for $x>0.4$. This spread near $x \approx 0.4$ is more pronounced for $\Sigma^{+}$ with lighter diquark than $\Xi^{o}$ carrying heavier diquark which clearly indicates the dominance of heavier diquark in carrying more momentum fraction $x$. However, $\Lambda$ being similar to $\Sigma$ portrays an alike trend of plots with comparatively little less spread and magnitude of the distortion which is because of its smaller mass. 

Comparative analysis of $s$ flavored quark in $\Lambda$, $\Sigma^{+}$ and $\Xi^{o}$ also shows an expansion on moving to $x=0.4$ but  beyond this a slow contraction for $\Sigma^{+}$ presents its ability to carry momentum fraction larger than $0.4$ which is because of the presence of $s$ flavor of quark and a lighter diquark in $\Sigma^{+}$. However,  as the diquark in $\Xi^{o}$ becomes heavier, the ability of carrying a momentum fraction larger than $0.4$ starts declining for an active $s$ flavored quark. In the sequence of slow contraction, $\Lambda$ lies in between $\Sigma^{+}$ and $\Xi^{o}$ as it also contains a lighter diquark just as $\Sigma$ but again small mass of $\Lambda$ baryon influences the positioning of it in this sequence.

In general, on comparing the transverse distortion between an active $u$ and $s$ flavored quark, one can observe a similar spread for momentum fraction for $x=0.4$ but a very slow shrink for $x>0.4$ in case of $s$ flavor of quark which implies that a massive flavored quark has a tendency of holding comparatively large momentum fraction than lighter flavored quark. 

\section{Single spin asymmetries \label{secSSA}}
\begin{figure*}
	\centering
	\begin{minipage}[c]{0.98\textwidth}
		\includegraphics[width=9 cm]{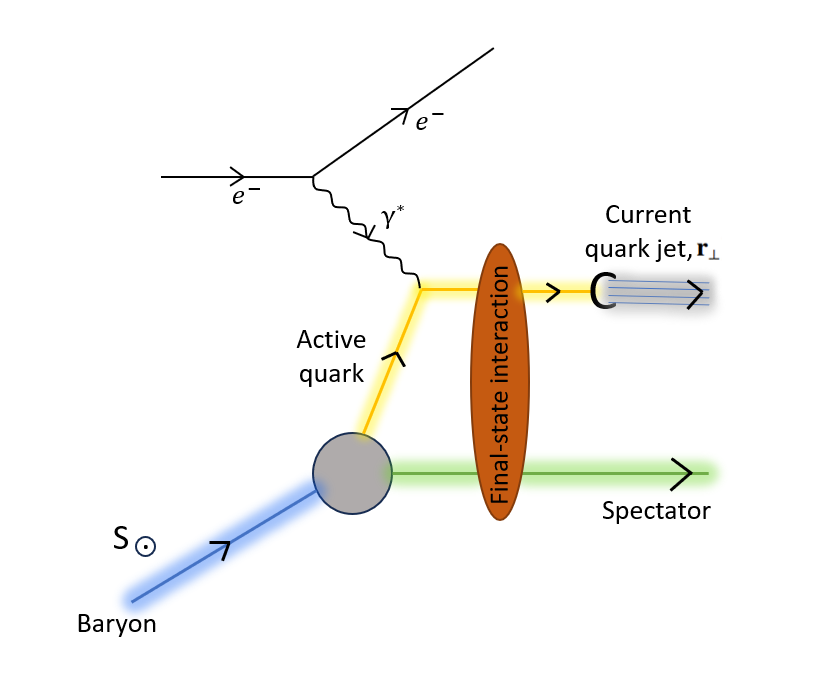}
		\hspace{0.03cm}
	\end{minipage}
	\caption{\label{FigFSI} (Color online) Final-state interaction in the semi-inclusive deep inelastic scattering.}
\end{figure*}

\begin{figure*}
	\centering
	\begin{minipage}[c]{0.98\textwidth}
		(a)\includegraphics[width=7cm]{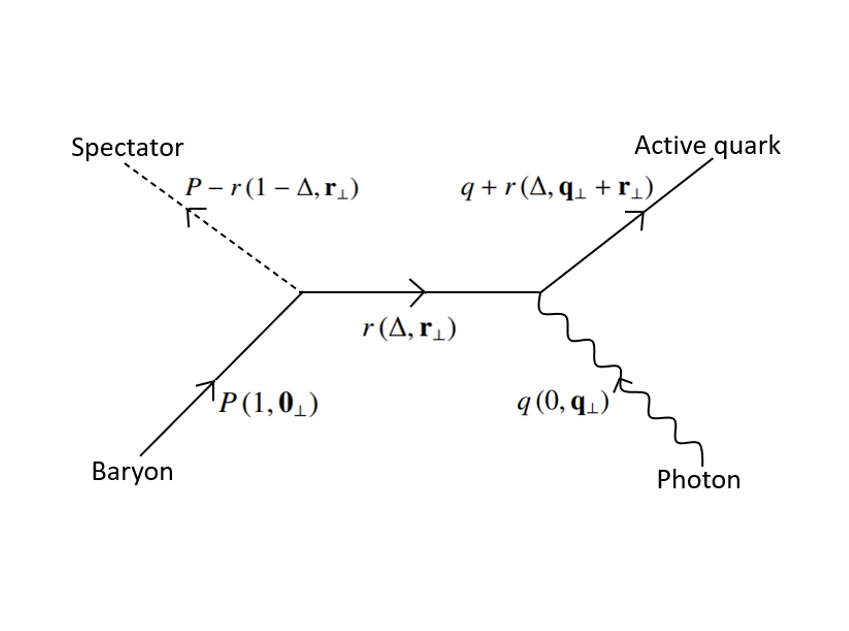}
		\hspace{0.03cm}
		(b)\includegraphics[width=7cm]{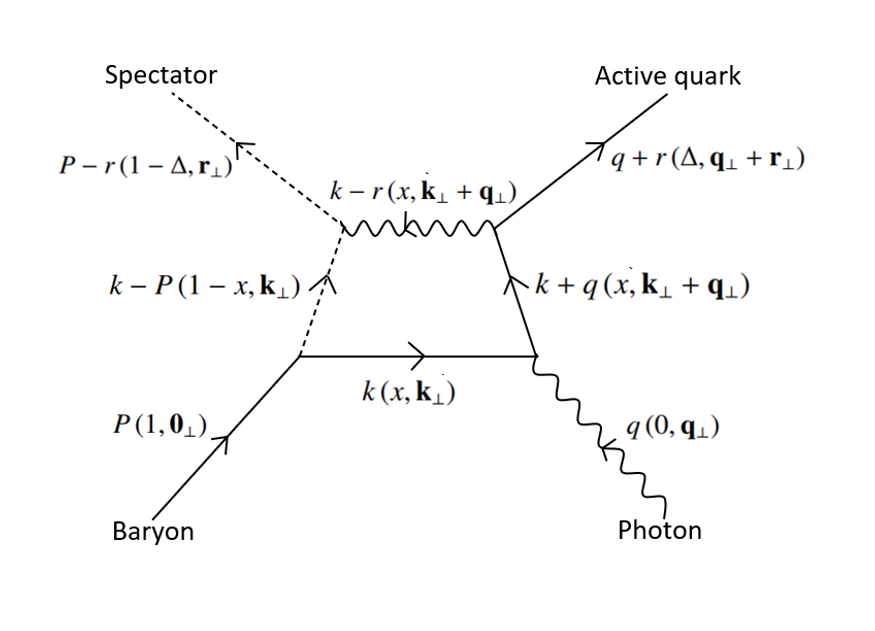}
		\hspace{0.03cm} 
	\end{minipage}
	\caption{\label{FigFeyn} (Color online) Tree and one-loop Feynman diagrams for $X+\gamma^{\ast} \rightarrow q + \mu_{n}^0$. } 
\end{figure*}

FSI between an active quark that has been already influenced by the virtual photon and its unaffected spectator is demonstrated in Fig. \ref{FigFSI}. To produce single-spin asymmetry, the required phase can be incorporated by the inclusion of this FSI and interference among the amplitudes of a process $X+\gamma^{\ast} \rightarrow q + \mu_{n}^0$, the superscript on the diquark mass corresponds to zero spin. Fig. \ref{FigFeyn} represents the tree and one-loop Feynman graphs that can be employed to obtain amplitude of the mentioned process. The structure of amplitudes \cite{Brodsky:2002cx}
\be
\mathcal{A}_{X}^{q}(\Uparrow \rightarrow \uparrow)&=&\bigg(M_{X}+\frac{m_{q}}{\Delta}\bigg) \, C \bigg(h_{X} +i \frac{e_{1}e_{2}}{8 \pi} I_{1X}\bigg) \, , \label{Amp1} \\
\mathcal{A}_{X}^{q}(\Downarrow \rightarrow \uparrow)&=&\bigg(\frac{r_{1}-i \, r_{2}}{\Delta}\bigg) \, C \bigg(h_{X} +i \frac{e_{1}e_{2}}{8 \pi} I_{2X}\bigg) \,, \label{Amp2} \\
\mathcal{A}_{X}^{q}(\Uparrow \rightarrow \downarrow)&=&\bigg(-\frac{r_{1}+i \, r_{2}}{\Delta}\bigg) \, C \bigg(h_{X} +i \frac{e_{1}e_{2}}{8 \pi} I_{2X}\bigg) \,, \label{Amp3} \\
\mathcal{A}_{X}^{q}(\Downarrow \rightarrow \downarrow) &=& \bigg(M_{X} + \frac{m_{q}}{\Delta}\bigg) \, C \bigg(h_{X} +i \frac{e_{1}e_{2}}{8 \pi} I_{1X}\bigg) \, , \label{Amp4}
\ee
with
\be
C&=&-g \, e_{1} P^{+} \sqrt{\Delta} \, 2 \, \Delta (1-\Delta) \, , \\
h_{X}&=&\frac{1}{\bfr+\mathcal{M}_{X}^{un}(\Delta)} \, ,\\
I_{1X}&=&\int_{0}^{1} d\alpha \frac{1}{\alpha (1-\alpha)\bfr^{2} + \alpha \mu_{g}^{2}+(1-\alpha) \mathcal{M}_{X}^{un}(\Delta)} \, , \\
I_{2X}&=&\int_{0}^{1} d\alpha \frac{\alpha}{\alpha (1-\alpha)\bfr^{2} + \alpha \mu_{g}^{2}+(1-\alpha) \mathcal{M}_{X}^{un}(\Delta)} \, .
\ee
Here, $e_{1}$ and $e_{2}$ are the electric charges of an active flavor of quark and a scalar spectator diquark respectively with $g$ as a coupling constant of the baryon- active quark- scalar spectator vertex. In the structure of amplitudes, the first term is a result of the Born contribution of tree representation whereas the second term corresponds to the one-loop representation that contains two different contributions $I_{1X}$ and $I_{2X}$. Their difference is infrared finite which can be regulated by the gauge particle mass $\mu_{g}$. In the Bjorken scaling limit, when $Q^{2}$ and laboratory energy of the photon $\nu$ is large with fixed $\Delta=x_{bj}$, the light-cone laboratory frame and usual laboratory frame becomes alike.

We have availed the light-cone gauge, $A^{+}=0$ which reduces the gauge link associated with an active quark (Wilson line) to unity. Hence, an active outgoing quark does not experience any FSI.  This paradox was unraveled in Ref. \cite{Brodsky:2005yw} by assuming the production of a hidden strange pair by gluon cleaving. As per the requirement of the Wilson line, the struck $s$ quark interacts in the final state through gluon exchange. For the four one-loop amplitudes, the covariant expressions can be jotted down as
\be
\mathcal{A}^{one-loop}_{X(q)}(I) &=& i \, g \, e_{1}^{2} e_{2} \int \frac{d^{4}k}{(2\pi)^4} \nonumber \\ &\times& \frac{\mathcal{N}_{X}^{q}(I)}{(k^{2}-m_{q}^{2}+i\epsilon)((k+q)^{2}-m_{q}^{2}+i\epsilon) ((k-r)^{2}-\mu_{g}^{2}+i\epsilon)((k-P)^{2}-\mu_{n}^{2}+i\epsilon)}\, .
\label{AmpOneLoop1}
\ee
From the Feynman diagram of these interaction, the numerator $\mathcal{N(I)}$ can be obtained as
\be
\mathcal{N}_{X}^{q}(\Uparrow \rightarrow \uparrow)&=& 2 P^{+} \sqrt{\Delta} \,  x \, \bigg(M_{X}+\frac{m_{q}}{x}\bigg)  \, q^{-} \, , \label{NAmp1} \\
\mathcal{N}_{X}^{q}(\Downarrow \rightarrow \uparrow)&=& 2 P^{+} \sqrt{\Delta} \, x \, (k^{1}-i k^{2}) \, q^{-}  \,, \label{NAmp2} \\
\mathcal{N}_{X}^{q}(\Uparrow \rightarrow \downarrow)&=& -2 P^{+} \sqrt{\Delta} \, x \,  (k^{1}+i k^{2}) \, q^{-} \,, \label{NAmp3} \\
\mathcal{N}_{X}^{q}(\Downarrow \rightarrow \downarrow) &=& 2 P^{+} \sqrt{\Delta} \, x \, \bigg(M_{X}+\frac{m_{q}}{x}\bigg)  \, q^{-} \, , \label{NAmp4}
\ee
where $x=\frac{k^{+}}{P^{+}}$ and $q^{-}=\frac{Q^{2}}{\Delta P^{+}}$. The term $q^{-}$ arises from the contribution of the gauge propagator in the Feynman diagram. The integration of Eq. (\ref{AmpOneLoop1}) gives a non-zero value for $x \in [0,1]$ and in the $\Delta<x<1$ region we have
\be
\mathcal{A}^{one-loop}_{X(q)}(I) &=& i \, g \, e_{1}^{2} e_{2} \, (2 \pi i) \int \frac{d^{2}\bfk}{2(2\pi)^4} \int P^{+} dx \frac{\mathcal{N}_{X}^{q}(I)}{P^{+4} x \, x (x-\Delta)(1-x)} \nonumber \\ 
&\times& \frac{1}{\bigg(P^{-}-\frac{(\mu_{n}^{2}+\bfk^{2})-i \epsilon}{(1-x)P^{+}}-\frac{(m_{q}^{2}+\bfk^{2})-i \epsilon}{x \, P^{+}}\bigg) \bigg(P^{-}-\frac{(\mu_{n}^{2}+\bfk^{2})-i \epsilon}{(1-x)P^{+}}+q^{-}-\frac{(m_{q}^{2}+(\bfk+\bfq)^{2})-i \epsilon}{x \, P^{+}}\bigg)} \nonumber \\ 
&\times& \frac{1}{\bigg(P^{-}-\frac{(\mu_{n}^{2}+\bfk^{2})-i \epsilon}{(1-x)P^{+}}+r^{-}-\frac{(\mu_{g}^{2}+(\bfk+\bfq)^{2})-i \epsilon}{x \, P^{+}}\bigg)}.
\label{AmpOneLoop2}
\ee
\begin{figure*}
	\centering
	\begin{minipage}[c]{0.98\textwidth}
		(a)\includegraphics[width=7cm]{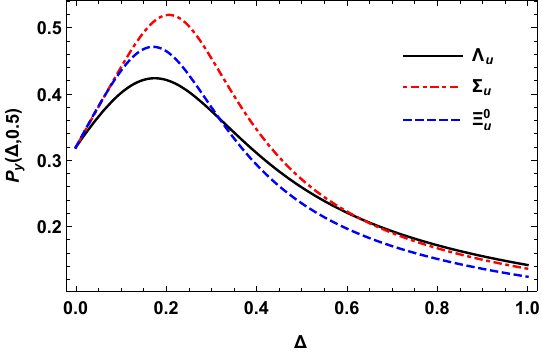}
		\hspace{0.03cm}
		(b)\includegraphics[width=7cm]{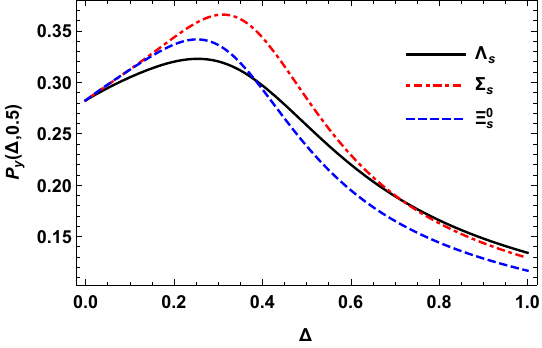}
		\hspace{0.03cm} 
	\end{minipage}
	\caption{\label{Fig7SSAd} (Color online) Juxtapose of azimuthal single-spin asymmetry as a function of light-front momentum fraction $\Delta$ for for $(a)\,u$ and $(b)\,s$ flavor of quarks in $\Lambda,\, \Sigma$ and $\, \Xi^{o}$.} 
\end{figure*}
The obtained result is identical to the light-cone time-ordered perturbation theory. The phase required to get single-spin asymmetries arrives from the imaginary part of Eq. (\ref{AmpOneLoop2}) that emerges from the potentially real intermediate state allowed before the re-scattering. The imaginary part of the propagator gives 
\be
-i \, \pi \, \delta \, \bigg(P^{-}-\frac{(\mu^{2}_{n}+\bfk^{2})}{(1-x) P^{+}}+q^{-}-\frac{(m_{q}^{2}+(\bfk+\bfq)^{2})-i \epsilon}{x \, P^{+}}\bigg) =-i \, \pi \frac{1}{P^{+}} \frac{\Delta^{2}}{\bfq^{2}} \, \delta(x-\Delta-\bar{\delta}) \, , 
\ee
where
\be
\bar{\delta}=2 \, \Delta \frac{\bfq \cdot (\bfk-\bfr)}{\bfq^{2}} \, . 
\ee
Since the exchanged momentum is small, the light-cone energy denominator for a gauge boson propagator gets dominance due to the presence of $\frac{(\bfk^{2}-\bfr^{2})+\mu_{g}^{2}}{(x-\delta)}$ term. The presence of one more $(x-\delta)$ term in the denominator gives only $(\bfk^{2}-\bfr^{2})+\mu_{g}^{2}$ which is independent of whether the photon is absorbed or emitted. The yield from the $0<x<\Delta$ region offers compliment to the yield of $\Delta<x<1$. The single-spin asymmetry can be defined in terms of the amplitudes as
\be
\mathcal{P}_{y}^{X(q)}=&\frac{1}{\mathcal{C}}& \big(i \, (\mathcal{A}_{X}^{q}(\Uparrow \rightarrow \uparrow)^{\ast} \, \mathcal{A}_{X}^{q}(\Downarrow \rightarrow \uparrow)-\mathcal{A}_{X}^{q}(\Uparrow \rightarrow \uparrow) \, \mathcal{A}_{X}^{q}(\Downarrow \rightarrow \uparrow)^{\ast}) \nonumber \\
&+& i \, (\mathcal{A}_{X}^{q}(\Uparrow \rightarrow \downarrow)^{\ast} \, \mathcal{A}_{X}^{q}(\Downarrow \rightarrow \downarrow)-\mathcal{A}_{X}^{q}(\Uparrow \rightarrow \downarrow) \, \mathcal{A}_{X}^{q}(\Downarrow \rightarrow \downarrow)^{\ast}) \big) \, ,
\ee
where the normalization $\mathcal{C}$ for the unpolarized cross section is
\be
\mathcal{C}=|\mathcal{A}_{X}^{q}(\Uparrow \rightarrow \uparrow)|^{2} + |\mathcal{A}_{X}^{q}(\Downarrow \rightarrow \uparrow)|^{2} +|\mathcal{A}_{X}^{q}(\Uparrow \rightarrow \downarrow)|^{2} +|\mathcal{A}_{X}^{q}(\Downarrow \rightarrow \downarrow)|^{2} .
\ee
\begin{figure*}
	\centering
	\begin{minipage}[c]{0.98\textwidth}
		(a)\includegraphics[width=7cm]{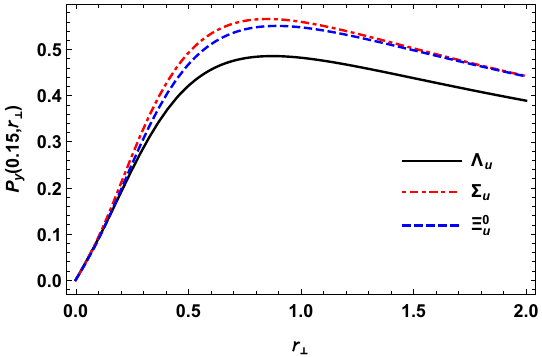}
		\hspace{0.03cm}
		(b)\includegraphics[width=7cm]{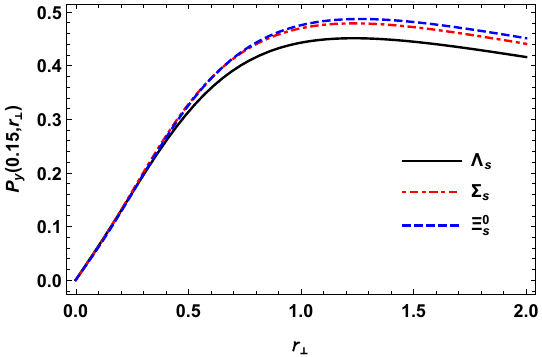}
		\hspace{0.03cm} 
	\end{minipage}
	\caption{\label{Fig8SSAr} (Color online) Juxtapose of azimuthal single-spin asymmetry as a function of $|r_{\perp}|$ for for $(a)\,u$ and $(b)\,s$ flavor of quarks in $\Lambda,\, \Sigma$ and $\, \Xi^{o}$.} 
\end{figure*}
For convenience, it was assumed that FSI produce a phase when exponentiated and the rescattering phases $e^{i \, \mathcal{X}_{i}} \, (i=1,2)$ with $\mathcal{X}_{i}=tan^{-1}(\frac{e_{1} e_{2}}{8 \pi} \frac{g_{i}}{h})$ are different for the spin-parallel and spin-antiparallel amplitudes. The phase difference in the spin-flip amplitudes originates from the orbital angular momentum $\bfk$ factor and the difference $\mathcal{X}_{1}-\mathcal{X}_{2}$ which is the cause of SSA is infrared finite. The azimuthal SSA transverse to the production plane $\hat{z}-\hat{x}$, defined by virtual photon and produced baryon is given by
\be 
\mathcal{P}_{y}^{X(q)} &=& -\frac{e_{1} e_{2}}{8 \pi} \, \frac{2 \, \left(\Delta \, M_{X}+m_{q}\right) \, r_{1}}{\left(\Delta \, M_{X}+m_{q}\right)^{2}+\bfr^{2}} \, \left[\bfr^{2}+\mathcal{M}^{un}_{X}(\Delta)\right] \, \frac{1}{\bfr^{2}} \, \ln\frac{\bfr^{2}+\mathcal{M}^{un}_{X}(\Delta)}{\mathcal{M}^{un}_{X}(\Delta)}.
\label{SSA}
\ee 
The presence of linear factor $r^{1}$ in the above Eq. (\ref{SSA}) assures the fact that SSA is proportional to the $S_{X} \cdot (\bfq \times \bfr)$ term. FSI from gluon exchange has the strength of $\frac{e_{1} e_{2}}{4 \pi} \rightarrow C_{F} \ \alpha_{s}(\mu^{2})$. The transferred momentum carried by the gluon $\mu^{2}=\frac{(\bfk-\bfr)^{2}}{e^{5/3}}$ has been used to to set the scale of $\alpha_{s}$ in the $\overline{MS}$ scheme. 

The dependence of SSA on the quark light-front momentum fraction $\Delta$ and the magnitude of an active quark momentum jet $\bfr$ relative to the virtual photon direction has been pictorially represented for composite quark flavors of stranged baryons in Figs. \ref{Fig7SSAd} and \ref{Fig8SSAr}  respectively. We observe that the SSA is recognizable at smaller values of quark light front momentum fraction implying that, in this region, correlation of the targeted baryon spin and the virtual photon is more effective to the hadron production plane. For a fixed value of momentum carried by an outgoing active quark i.e. $r_{\perp}=0.15$, comparative demonstration of SSA for $u$ quark in $\Lambda, \Sigma$ and $\Xi^{o}$ baryons shows that $\Sigma$ with lighter diquark has the most SSA, gains a peak at $\Delta=0.206$ and then decreases slowly. However, peak comparatively at smaller value of $\Delta=0.171$ and fast decrement has been observed for $\Xi^{o}$ baryon due to the dominance of a heavy diquark. For $u$ flavord quark in $\Lambda$, a decrement in the amplitude, shifting of a peak towards smaller value of $\Delta=0.174$ and slow reduction with $\Delta$ are the key attributes of the smaller mass of $\Lambda$ baryon because beside the difference of baryon masses, $\Lambda$ and $\Sigma$ have similar contribution of diquark. The logarithmic term contributes the most in the variation of SSA for different baryons. Similar trend for an active $s$ flavored quark can be observed which contains $\Sigma$ baryon on the top of the list with the maximum amplitude and a peak at $\Delta=0.311$. Next, there is a $\Xi^{o}$ baryon for which amplitude is reduced with the shifting of peak towards smaller value of $\Delta$ at $0.252$ due to the effect of comparatively heavier diquark $us$ than $uu$ in $\Sigma$. At last, we have a $\Lambda$ baryon with the smallest amplitude and peak at $\Delta=0.254$ as a result of the mass of a $\Lambda$ baryon. 
\begin{figure*}
	\centering
	\begin{minipage}[c]{0.98\textwidth}
		(a)\includegraphics[width=7cm]{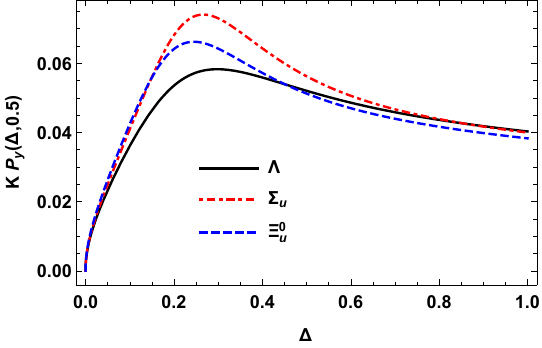}
		\hspace{0.03cm}
		(b)\includegraphics[width=7cm]{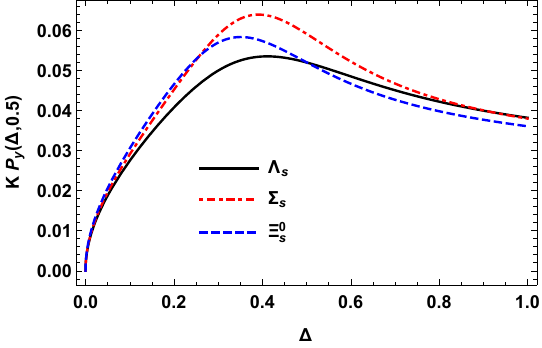}
		\hspace{0.03cm} 
	\end{minipage}
	\caption{\label{Fig9KSSAd} (Color online) Juxtapose of azimuthal single-spin asymmetry as a function of light-front momentum fraction $\Delta$ for for $(a)\,u$ and $(b)\,s$ flavor of quarks in $\Lambda,\, \Sigma$ and $\, \Xi^{o}$.}
\end{figure*}
A comparison between $u$ and $s$ flavored quark of same baryons suggests that $s$ quark flavor can carry more light-front momentum fraction as the peaks for $s$ flavored quarks are at higher value of $\Delta$ than $u$ flavored quarks however, with a spread over large region of light-front momentum fraction and a lower chance of having SSA as their amplitude is less than $u$ flavored quarks. 

Fig. \ref{Fig8SSAr}, representing the dependence of SSA on the magnitude of an active quark momentum jet $\bfr$ relative to the virtual photon direction at $\Delta=0.15$, shows that for an outgoing $u$ flavored quark, SSA in $\Sigma$ shows minimal difference as compared to SSA in  $\Xi^{o}$ but there is a significant difference from the value for $\Lambda$. Such difference may be arising due to the dominance of diquark mass in $\Xi^{o}$ and mass of a baryon in $\Lambda$. However, for an outgoing $s$ quark, there is a small distinction among all three baryons with less amplitude of SSA and queued according to the decreasing order of the masses of baryons. Comparison between $u$ and $s$ quark shows that active $u$ quark momentum jet have more SSA with respect to $\bfr$ and significant difference lies only for $\Lambda$ baryon.

\begin{figure*}
	\centering
	\begin{minipage}[c]{0.98\textwidth}
		(a)\includegraphics[width=7cm]{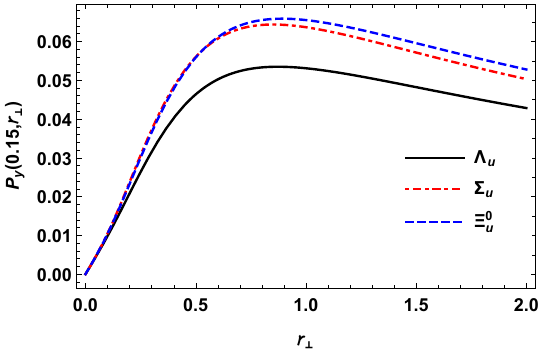}
		\hspace{0.03cm}
		(b)\includegraphics[width=7cm]{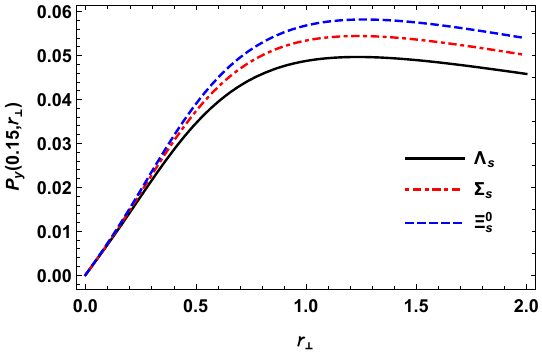}
		\hspace{0.03cm} 
	\end{minipage}
	\caption{\label{Fig10KSSAr} (Color online) Juxtapose of azimuthal single-spin asymmetry as a function of $|r_{\perp}|$ for $(a)\,u$ and $(b)\,s$ flavor of quarks in $\Lambda,\, \Sigma$ and $\, \Xi^{o}$.} 
\end{figure*}
The transverse azimuthal spin asymmetry $\mathcal{P}_y$ can be compared with experimentally measured HERMES transverse asymmetry $A^{sin \, \phi}_{UT}$ in which the polarization of the targeted baryon transverse to the incident lepton direction \cite{Brodsky:2004hh}. The longitudinal asymmetry $A^{sin \, \phi}_{UL}$ which corresponds to the targeted baryon, polarized along the direction of an incident lepton direction can be extracted from the transverse asymmetry by introducing a kinematical factor 
\be
K=\frac{Q}{\nu} \sqrt{1-y} \, = \, \sqrt{\frac{2 \, x \, M_{X}}{E}} \sqrt{\frac{1-y}{y}}, 
\ee
and the longitudinal asymmetry can be can be obtained as 
\be 
A^{sin \, \phi}_{UL} \, = \, K \, A^{sin \, \phi}_{UT}.
\ee
The resulting predictions of the longitudinal asymmetry $A^{sin \, \phi}_{UL}$ for strange baryons are presented in Figs. \ref{Fig9KSSAd} and \ref{Fig10KSSAr} as a function of $\Delta$ and $\bfr$ respectively for the constituent flavor of quarks. We have used the input for $E$ as $27.6 \, GeV$ \cite{HERMES:1998mat, HERMES:2020ifk}. The queue of the strange baryons and the trend of peak positioning of the longitudinal SSA as a function of $\Delta$ is same as in the case of transverse SSA. The more ability of a massive quark to carry comparatively larger momentum fraction $\Delta$ is reassured here. Fig. \ref{Fig10KSSAr} illustrates the decrement of the amplitude of longitudinal SSA as the mass of a baryon increases and this mass effect  attributes to the presence of a baryon mass term in kinematical factor $K$. The decrement in the portray of $K$  $\mathcal{P}_{y}$ at large values of $\bfr$ has been observed and the pace of this decrement is very slow for $s$ quark due to the presence of lighter diquark implying the tendency of massive $s$ flavor quark to carry large quark momentum.
\section{Conclusion \label{secCon}}
We have presented the mapping of transverse distortion and single-spin asymmetry of strange octet baryons carrying same spin-parity quantum number. Their spatial mappings have been demonstrated by using the QCD inspired quark-scalar diquark model and the mass effect of the parent baryon and constituent quarks on the deformation has also been studied. The presence of finite spin-flip quantum fluctuations of GPD is a measure of the transverse deformation in an active quark distribution. These transverse distortions show that a massive quark undergoes left-right asymmetry to greater extent and has a tendency of carrying more longitudinal momentum fraction than a light quark as the distribution undergoes a slow contraction for high longitudinal momentum fraction. 

The dependence of single-spin asymmetry on quark light-front momentum fraction reassures the capacity of a massive quark of carrying significantly high longitudinal momentum fraction but with less likeliness of single-spin asymmetry. On the other hand, the difference in the mass of diquark is not proven effective to produce a difference in single-spin asymmetry with the magnitude of momentum carried by an outgoing quark but the tendency of having single-spin asymmetry decrease with decrement of a baryon mass. Further, to relate these theoretical predictions, we demonstrate the azimuthal longitudinal asymmetry that is accessible in scattering processes in HERMES and JLab experiment projects. JLab has a project over beam spin asymmetry and  proposed the measurements of azimuthal modulations in SIDIS for different types of hadron targets, and polarizations over a broad kinematic range \cite{Achenbach:2023pba} which can act as a baseline to study the asymmetries in baryons. \par
\section{Acknowledgement \label{secAck}}
H.D. would like to thank  the Science and Engineering Research Board, Department of Science
and Technology, Government of India through the grant (Ref No.TAR/2021/000157) under TARE
scheme for financial support.


\begin{thebibliography}{100}
\section*{References}
\bibitem{Ji:2021qgo}
X.~Ji, Y.~Liu and A.~Sch\"afer,
Nucl. Phys. B \textbf{971}, 115537 (2021).

\bibitem{Lorce:2021xku}
C.~Lorc\'e, A.~Metz, B.~Pasquini and S.~Rodini,
JHEP \textbf{11}, 121 (2021).

\bibitem{Wang:2019mza}
R.~Wang, J.~Evslin and X.~Chen,
Eur. Phys. J. C \textbf{80}, 507 (2020).

\bibitem{Lorce:2017xzd}
C.~Lorc\'e,
Eur. Phys. J. C \textbf{78}, 120 (2018).

\bibitem{Ji:2016jgn}
X.~Ji, F.~Yuan and Y.~Zhao,
Phys. Rev. Lett. \textbf{118}, 192004 (2017).

\bibitem{Alexandrou:2020sml}
C.~Alexandrou, S.~Bacchio, M.~Constantinou, J.~Finkenrath, K.~Hadjiyiannakou, K.~Jansen, G.~Koutsou, H.~Panagopoulos and G.~Spanoudes,
Phys. Rev. D \textbf{101}, 094513 (2020).

\bibitem{Cruz-Torres:2019fum}
R.~Cruz-Torres, D.~Lonardoni, R.~Weiss, N.~Barnea, D.~W.~Higinbotham, E.~Piasetzky, A.~Schmidt, L.~B.~Weinstein, R.~B.~Wiringa and O.~Hen,
Nature Phys. \textbf{17}, 306-310 (2021).

\bibitem{West:2020tyo}
J.~R.~West,
Nucl. Phys. A \textbf{1029}, 122563 (2023).

\bibitem{Celiberto:2021zww}
F.~G.~Celiberto,
Nuovo Cim. C \textbf{44}, 36 (2021).

\bibitem{Burkardt:2006td}
M.~Burkardt,
AIP Conf. Proc. \textbf{915}, 313-318 (2007).

\bibitem{Bacchetta:2022crh}
A.~Bacchetta, F.~G.~Celiberto and M.~Radici,
Rev. Mex. Fis. Suppl. \textbf{3}, 0308108 (2022).

\bibitem{Barry:2023qqh}
P.~C.~Barry \textit{et al.} [Jefferson Lab Angular Momentum (JAM)],
Phys. Rev. D \textbf{108},L091504 (2023).

\bibitem{Meissner:2009ww}
S.~Meissner, A.~Metz and M.~Schlegel,
JHEP \textbf{08}, 056 (2009).

\bibitem{Maji:2016yqo}
T.~Maji and D.~Chakrabarti,
Phys. Rev. D \textbf{94}, 094020 (2016).

\bibitem{Diehl:2015uka}
M.~Diehl,
Eur. Phys. J. A \textbf{52},149 (2016).

\bibitem{Boffi:2007yc}
S.~Boffi and B.~Pasquini,
Riv. Nuovo Cim. \textbf{30}, 387-448 (2007).

\bibitem{Pasquini:2008ax}
B.~Pasquini, S.~Cazzaniga and S.~Boffi,
Phys. Rev. D \textbf{78}, 034025 (2008).

\bibitem{Xie:2023xkz}
G.~Xie, W.~Kou, Q.~Fu, Z.~Ye and X.~Chen,
Eur. Phys. J. C \textbf{83}, 900 (2023).

\bibitem{Goloskokov:2007nt}
S.~V.~Goloskokov and P.~Kroll,
Eur. Phys. J. C \textbf{53}, 367-384 (2008).

\bibitem{Brooks:2018uqk}
W.~Brooks, I.~Schmidt and M.~Siddikov,
Phys. Rev. D \textbf{98}, 116006 (2018).

\bibitem{Bacchetta:2010uj}
A.~Bacchetta,
AIP Conf. Proc. \textbf{1374}, 29-34 (2011).

\bibitem{Collins:2004nx}
J.~C.~Collins and A.~Metz,
Phys. Rev. Lett. \textbf{93}, 252001 (2004).

\bibitem{Accardi:2017pmi}
A.~Accardi and A.~Bacchetta,
Phys. Lett. B \textbf{773}, 632-638 (2017).

\bibitem{Belitsky:2005qn}
A.~V.~Belitsky and A.~V.~Radyushkin,
Phys. Rept. \textbf{418}, 1-387 (2005).

\bibitem{Kumar:2015fta}
N.~Kumar and H.~Dahiya,
Int. J. Mod. Phys. A \textbf{30}, 1550010 (2015).

\bibitem{Kaur:2018ewq}
N.~Kaur, N.~Kumar, C.~Mondal and H.~Dahiya,
Nucl. Phys. B \textbf{934}, 80-95 (2018).

\bibitem{Zhang:2016qqg}
J.~Zhang and B.~Q.~Ma,
Phys. Rev. C \textbf{93}, 065209 (2016).

\bibitem{Diehl:2003ny}
M.~Diehl,
Phys. Rept. \textbf{388}, 41-277 (2003).

\bibitem{Kim:2008ghb}
D.~S.~Kim, D.~S.~Hwang and J.~Kim,
Phys. Lett. B \textbf{669}, 345-351 (2008).

\bibitem{Kumar:2014coa}
N.~Kumar and H.~Dahiya,
Phys. Rev. D \textbf{90}, 094030 (2014).

\bibitem{Burkardt:2002hr}
M.~Burkardt,
Int. J. Mod. Phys. A \textbf{18}, 173-208 (2003).

\bibitem{Burkardt:2001ni}
M.~Burkardt,
doi:10.1142/9789812799708\_0006
[arXiv:hep-ph/0105324 [hep-ph]].

\bibitem{Burkardt:2003je}
M.~Burkardt and D.~S.~Hwang,
Phys. Rev. D \textbf{69}, 074032 (2004).

\bibitem{HERMES:1999ryv}
A.~Airapetian \textit{et al.} [HERMES],
Phys. Rev. Lett. \textbf{84}, 4047-4051 (2000).

\bibitem{Bravar:1999rq}
A.~Bravar,
Nucl. Phys. B Proc. Suppl. \textbf{79}, 520-522 (1999).

\bibitem{Boer:2002ju}
D.~Boer, S.~J.~Brodsky and D.~S.~Hwang,
Phys. Rev. D \textbf{67}, 054003 (2003).

\bibitem{Brodsky:2002pr}
S.~J.~Brodsky, D.~S.~Hwang and I.~Schmidt,
Phys. Lett. B \textbf{553}, 223-228 (2003).

\bibitem{Brodsky:2002cx}
S.~J.~Brodsky, D.~S.~Hwang and I.~Schmidt,
Phys. Lett. B \textbf{530}, 99-107 (2002).

\bibitem{Mukherjee:2002xi}
A.~Mukherjee and M.~Vanderhaeghen,
Phys. Rev. D \textbf{67}, 085020 (2003).

\bibitem{Dodson:2021rdq}
J.~Dodson, S.~Bhattacharya, K.~Cichy, M.~Constantinou, A.~Metz, A.~Scapellato and F.~Steffens,
PoS \textbf{LATTICE2021}, 054 (2022).

\bibitem{sstwist4}
S.~Sharma and H.~Dahiya,
Int. J. Mod. Phys. A \textbf{37}, 2250205 (2022).

\bibitem{Hwang:2007tb}
D.~S.~Hwang and D.~Mueller,
Phys. Lett. B \textbf{660}, 350-359 (2008).

\bibitem{Lorce:2017wkb}
C.~Lorc\'e, L.~Mantovani and B.~Pasquini,
Phys. Lett. B \textbf{776}, 38-47 (2018).

\bibitem{Brodsky:2005yw}
S.~J.~Brodsky,
SLAC-PUB-11568 (2005).


\bibitem{Dirac:1949cp}
P.~A.~M.~Dirac,
Rev. Mod. Phys. \textbf{21}, 392-399 (1949).

\bibitem{Brodsky:2000rt}
S.~J.~Brodsky,
Nucl. Phys. B Proc. Suppl. \textbf{90}, 3-13 (2000).

\bibitem{Brodsky:2000ii}
S.~J.~Brodsky, D.~S.~Hwang, B.~Q.~Ma and I.~Schmidt,
Nucl. Phys. B \textbf{593}, 311-335 (2001).

\bibitem{Brodsky:2000xy}
S.~J.~Brodsky, M.~Diehl and D.~S.~Hwang,
Nucl. Phys. B \textbf{596}, 99-124 (2001).


\bibitem{Brodsky:2004hh}
S.~J.~Brodsky,
Acta Phys. Polon. B \textbf{36}, 635-656 (2005).

\bibitem{HERMES:1998mat}
K.~Ackerstaff \textit{et al.} [HERMES],
Nucl. Instrum. Meth. A \textbf{417}, 230-265 (1998).

\bibitem{HERMES:2020ifk}
A.~Airapetian \textit{et al.} [HERMES],
JHEP \textbf{12}, 010 (2020).

\bibitem{Achenbach:2023pba}
P.~Achenbach, 
[arXiv:2303.02579 [hep-ph]].
\end{thebibliography}
\end{document}